\documentclass[english,doublecol,a4paper]{epl2}
\usepackage[]{fontenc}
\usepackage[latin9]{inputenc}
\usepackage{color}
\usepackage{graphicx}
\usepackage{amssymb}

\makeatletter

\usepackage{array}

\usepackage{epsfig}

\usepackage{epstopdf}

\usepackage{color}

\usepackage{babel}
\makeatother

\begin{document}

\title{Influence of laser-excited electron distributions on the x-ray \\
magnetic
circular dichroism spectra: Implications for femtosecond demagnetization in Ni}

\shorttitle{Influence of laser-excited electron distributions on x-ray spectrum of Ni} %Insert here a short version of the title if it exceeds 70 characters

\author{K. Carva\thanks{E-mail: \email{karel.carva@fysik.uu.se}} \and D.
Legut \and P. M. Oppeneer}
\shortauthor{K. Carva \etal} 
\institute{ Department of Physics and Materials Science, Box 530, Uppsala
University, S-75121 Uppsala, Sweden}




\abstract{
In pump-probe experiments an intensive laser pulse creates 
non-equilibrium excited electron distributions in the first few hundred femtoseconds after the pulse.
The influence of non-equilibrium electron distributions caused by
a pump laser on the apparent X-ray magnetic circular dichroism (XMCD) signal of Ni is investigated
theoretically here for the first time, considering
electron distributions immediately after the pulse as well as thermalized
ones, that are not in equilibrium with the lattice or spin systems.
The XMCD signal is shown not to be simply proportional to the spin
momentum in these situations. The computed spectra are compared to recent
pump-probe XMCD experiments on Ni. 
We find that the majority of experimentally observed
features considered to be a proof of ultrafast spin momentum transfer to the lattice can 
alternatively be attributed to non-equilibrium electron distributions. Furthermore, we find the
 XMCD sum rules for the atomic spin and orbital magnetic moment to remain 
valid, even for the laser induced non-equilibrium electron distributions.  
%We deduce electron thermalization to occur in Ni at about 150 fs after ending of the pump pulse. }
}

\maketitle
\section{Introduction}
Ultrafast control of the magnetization of ferromagnetic materials
is essential for modern technologies such as high-speed magnetic recording
and spin electronics \cite{kimel05,stanciu07,berezovsky08}. Magneto-optical
pump-probe techniques have become exploited to investigate the relaxation
process and dynamics of the magnetization in the sub-picosecond regime
\cite{beaurepaire96,ju98,koopmans03}. In such experiments an intensive,
short laser pulse is applied to optically excite a material and subsequently
the magnetic response of the system is probed by a magneto-optical
technique, for example, the Kerr effect. Earlier pump-probe experiments
on Ni in the visible range showed a strong reduction of the Kerr signal
\cite{beaurepaire96}, yet, the mechanisms of the magnetization decay
were not well understood. The possible dissipation channels, the role
of the spin-orbit interaction (SOI), and the influence of the non-equilibrium
electron population have been discussed 
\cite{beaurepaire96,zhang00,koopmans00,regensburger00,kampfrath02}.

Stamm \textit{et al}. \cite{stamm07} have recently extended the pump-probe
technique by using the x-ray magnetic circular dichroism (XMCD)
to probe the magnetic response of a Ni film that was excited through
an intensive %780~nm 
pump laser pulse. The observed XMCD value at the Ni $L_{3}$ edge
was found to be strongly decreased, by 65\% of its original value,
which was interpreted as a quenching of the Ni magnetization within
120 fs.
As the XMCD technique gives access---through the XMCD sum rules---to
separate values for the spin and orbital magnetism, a new window on
their individual response is offered. 
Stamm \textit{et al.} \cite{stamm07} concluded from a sum rule application
that it is the spin component that becomes strongly reduced, whereas
the orbital moment is practically unchanged, and deduce accordingly
 that the spin becomes reduced within 120 fs through spin
 angular momentum transfer to the lattice.

However, there are several open questions. The sharp drop of the Kerr
signal observed in pump-probe experiments on Ni 
has been interpreted as evidence for demagnetization occurring through
SOI within 20 fs \cite{zhang00}, but it has also been attributed
to state-blocking effects 
\cite{koopmans00}, or to a combination of these \cite{gomez03}.
The state-blocking  effect occurs when certain unoccupied
states that are accessible to the probe laser under equilibrium conditions
are no longer accessible, as they have been populated already through
the pump laser, and also, some of the originally forbidden transitions
become allowed. 
Magneto-optical pump-probe experiments on Ni films \cite{koopmans00}
indicated that state-blocking effects
had disappeared from the transient Kerr response at about 250 fs after the beginning of the pump pulse.
%  On the other hand, 
More advanced magneto-optical pump-probe polarimetry experiments performed on CoPt$_3$ films
demonstrated that state-blocking effects had vanished from the magneto-optical response 150 fs after the onset of the pump \cite{r_02_Guido_CoPt3_MOValidity,r_04_Bigot_MOValidity}.
%have demonstrated that for times greater than 50fs after the pulse state-blocking effects
%do not affect MO response of CoPt$_3$ \cite{r_02_Guido_CoPt3_MOValidity,r_04_Bigot_MOValidity}.
The somewhat different relaxation times for CoPt$_3$ and Ni could be possible, because
%However 
CoPt$_3$ exhibits
%, among other features, 
a much stronger spin-orbit interaction than Ni and a very different band structure. Therefore, the relaxations
in these materials have to be considered separately.
%and one cannot come to conclusions about relaxation times for Ni from that data. }

Here we study computationally the influence of non-equilibrium 
effects on x-ray absorption and XMCD spectra. We show that such effects do 
considerably modify both the x-ray absorption spectrum (XAS) and XMCD spectrum, 
each in a specific way.
The predicted modification
of these spectra for the case of Ni is found to be in good agreement with
observed experimental features \cite{stamm07}.
We consider two forms of non-equilibrium
electron distributions, first, specific modifications of the population
of states due to the pump laser, which are expected to exist to about
100\,fs after the pulse, and second, \textit{thermalized} electron
distributions, which are expected at about 300--400\,fs after the pump
pulse \cite{hohlfeld00,delfatti00}. In the former case, the electron distribution
is not a Fermi-Dirac (FD) function, while in the latter case, the
thermalized electrons can be described by a FD distribution, but the
electrons are not yet in equilibrium with the lattice.

\section{Computational methodology} 
In our computational approach
we use, first, experimental information as well as
\textit{ab initio} calculated transition probabilities to determine
the amount of population or depopulation of Ni's spin-polarized band
states by the pump laser. These modified occupation numbers are subsequently
used in relativistic calculations of the 
XAS and XMCD spectrum, in which the exchange-splitting of the core
levels is included. The XAS and XMCD asymmetry are determined from
the dielectric tensor  %$\mathbf{\varepsilon}$,
%$\epsilon_{ij}$ ($i,\,j=x,y,$ or $z$),
whose elements we compute from the Kubo linear-response expression.
This computational scheme, combined with accurate relativistic band structure calculations, in which the exchange-splitting of the core states is taken into account, 
provides theoretical XAS and XMCD spectra that agree very
well with experiments \cite{kunes01,kunes03}.
%?
Our evaluations of the XAS and XMCD have been implemented on the 
%The calculations were done using 
full-potential, relativistic \texttt{WIEN2k} method \cite{wien2k},
%based on the relativistic full-potential linear augmented plane wave method,
%(FLAPW) 
employing the local spin-density approximation (LSDA). The full Brillouin
zone in the calculation comprised $10\:000$ $\mathbf{k}$ points,
the fcc Ni lattice parameter was 
$3.42\,\textrm{\AA}$. Lorentzian broadening of 
1.2 eV for the $L_{2}$ edge and 0.7 eV for the $L_{3}$ edge was
applied to the calculated spectra to represent the core-hole lifetime
and smoothen small features of spectra below the experimental resolution.
 We note that certain features of the electronic structure of Ni are not correctly described within the LSDA, in particular the width of the occupied $d$ band and the appearance of a photoemission satellite peak at 6 eV binding energy \cite{eastman80,r_77_Guill_Ni_PES_peak}. These discrepancies  can be corrected
 when on-site correlation effects are taken into account \cite{manghi97}.

The number of pump-laser excited electrons per atom of the studied
sample follows from the effective excitation density $f_{exc}$. Its
maximum value $f_{exc}^{max}$ reached due to the pulse can be estimated
from the information about the laser and the absorption in the material,
using the formula \cite{koopmans00}: %\[
$f_{exc}^{max}=$ $f_{abs}\left(d_{Ni}\right)\frac{E_{p}}{\hbar\omega}\frac{V_{Ni}}{Sd_{Ni}}\,,$
%\]
where $d_{Ni}$ is the thickness of the thin Ni film, $f_{abs}$ is
the fraction of incident light absorbed in the film, $V_{Ni}$ the
unit cell volume, $E_{p}$ the pulse energy, $\omega$ the frequency
of the pulse laser, 
and $S$ is the area of the spot irradiated by the pump laser. For
the experimental values given by Stamm \textit{et al}. 
we obtain the maximum excitation density $f_{exc}^{max}=0.76$,
where the fact that $15$\% of each laser pulse excites the sample
is taken into account, $d_{Ni}$=$15$ nm, $E_{p}$=$2$ mJ, $f_{abs}$=$0.1$,
 $\hbar \omega$=$1.59$ eV, and
$S$=$0.75$ mm$^{2}$. 
The fine accuracy of $f_{exc}^{max}$ estimate might be limited,
but what is important here is that 
it determines roughly the range of values for $f_{exc}$ to be
expected in the system.
 Our results are provided as a function of $f_{exc}$ for a wide range
of $f_{exc}$ starting from $0$.
We note that the present value of $f_{exc}^{max}$ is rather high;
the energy of the employed pump laser $E_{p}$ (2 mJ) is considerably
larger than, for example, that of $E_{p}$=$1.6$ nJ which was used
in an earlier pump-probe study of Ni \cite{koopmans00}. To avoid
very high occupations of bands above $E_{{\rm F}}$ we have restricted
our calculations to $f_{exc}$ values up to 0.6.

Next, we compute the occupation numbers of relevant bands in Ni, 
following the approach outlined earlier \cite{oppeneer04}. The simplest
picture of a non-FD 
distribution due to a pulse with fixed energy $\hbar\omega$ is the
following: occupied bands down to $\hbar\omega$ below $E_{\mathrm{F}}$
are partially depleted and previously unoccupied bands up to $\hbar\omega$
above $E_{\mathrm{F}}$ are partially populated. % (model A of Ref.\ \cite{oppeneer04}).
The occupation numbers of those bands that are accessible to the pump
laser are computed from \textit{ab initio} calculated optical intensities.
Note that the optical excitation conserves spin, thus we examine majority
and minority spin transitions separately. %new
The spin-resolved occupation of bands $f_{b,\sigma}^{>}$ above the
Fermi level is computed from $f_{b,\sigma}^{>}=\frac{w_{b,\sigma}w_{\sigma}}{n_{b,\sigma}}f_{exc}$,
where $w_{b,\sigma}$ is the spin-dependent weight of transitions
to band $b$, $w_{\sigma}$ the probability of transition with spin
$\sigma$, %=\sum_b w_{b,\sigma}$, 
and $n_{b,\sigma}$ is the total number of spin electrons that fits
in the band in the considered energy window\footnote{
Our \textit{ab initio} calculations give that close to $100$\% of
all majority-spin transitions go into valence band no.\ $11$, 
while the minority spin transitions are divided between bands no.\ $10$,
$11$, and $12$, with respective weights $w_{b,\downarrow}$ 52\%,
3\%, and 44\%. 
For $n_{b,\sigma}$, the number of electrons that fit in the relevant
bands, we obtain 0.39, 0.022, and 0.28 for spin-down electrons in
bands no.\ 10, 11, and 12, respectively, and 0.22 spin-up electrons
in band 11.%
}. 
Below $E_{\mathrm{F}}$ there are several flat bands with similar
transition strengths and a simpler approach is adopted here, neglecting
the small differences between band-resolved transition strengths,
but ensuring that spin-polarization of electrons removed from occupied
bands is correctly accounted.

After the initial laser excitation, fast electron-electron processes
cause electron equilibration on a time scale of about 
350 fs \cite{delfatti00}, leading
to a FD distribution with a well defined electron temperature $T_{e}$.
The electron temperature can be computed from the energy deposited
by the laser. % per atom. 
The average energy per excited Ni atom follows from $\langle{\cal {E}\rangle}$=
$\int Ef_{FD}(E,T)n(E)dE$, where $n(E)$ is the Ni density of states
(DOS). Using the computed Ni DOS, we obtain a maximal electron temperature
of 11\,300\,K, 
which corresponds to the laser excitation with $f_{exc}^{max}$. A
temporal evolution of the electron population is expected to reduce
$T_{e}$ already directly after the laser pulse. In our calculations
we therefore consider electron temperatures ranging from 1\,000 to
11\,000\,K. These temperatures are higher than the Curie temperature
of Ni ($T_{C}=630$ K), what is possible since the thermalized electrons
are not yet in equilibrium with the spins nor with the lattice. The
magnetic moment on Ni, however, computed for such a high-temperature
FD function, is reduced from the equilibrium moment, because there
exists a high spin-majority $3d$-DOS peak just below $E_{{\rm F}}$
whose contribution becomes diminished through the FD function.

%We note that the adopted approach neglects the change of bands 
%due to electron repopulation. This neglected contribution should lead to corrections
%of higher order in $f_{exc}$ than the effect of repopulation (mainly
%linear), which thus gives the trend dominant up to an unknown value
%of $f_{exc}$, where higher order effects would outweigh it.  

\section{Calculated results} 
\begin{figure}[tb]
\includegraphics[clip,width=0.95\columnwidth]{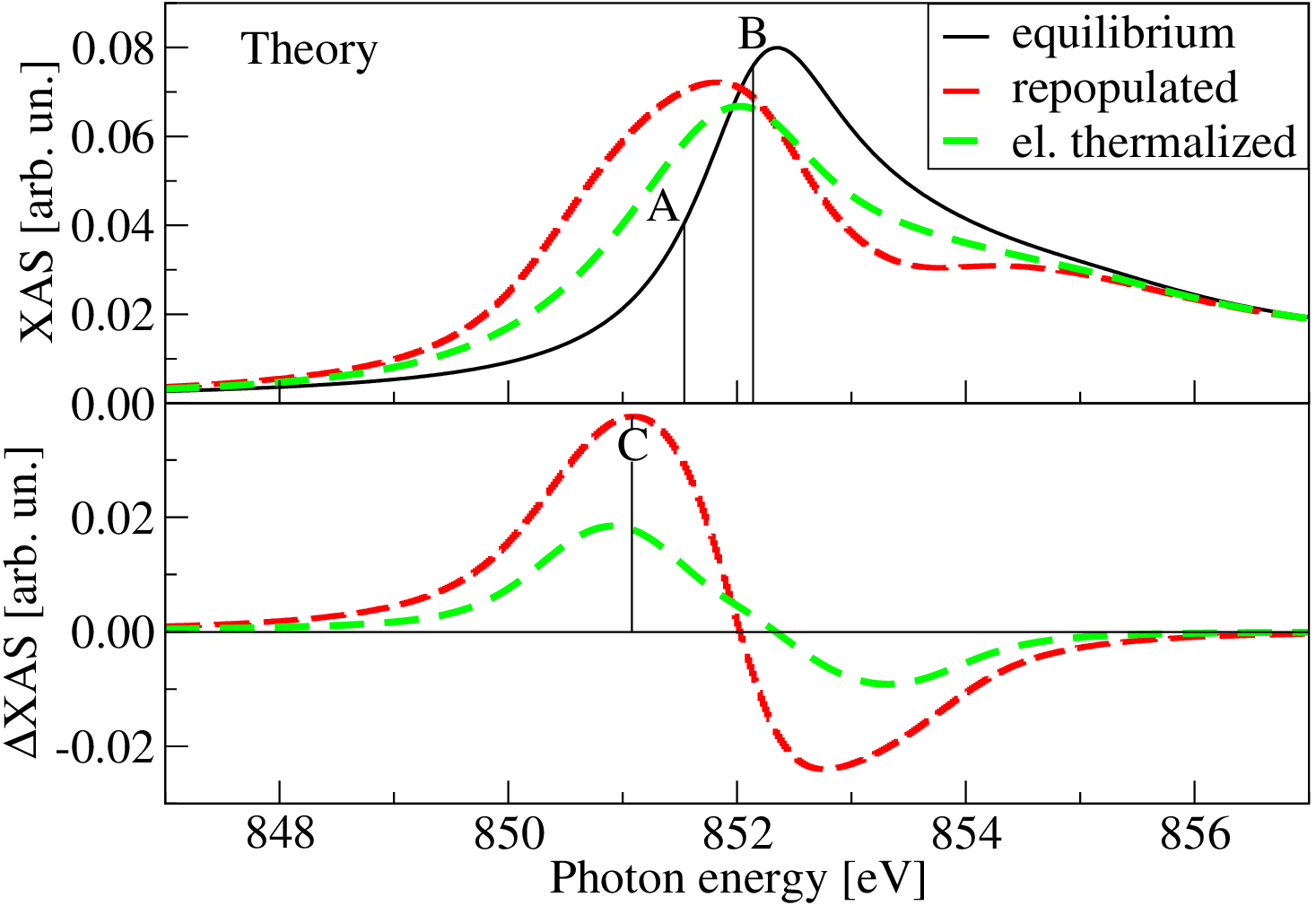}

\includegraphics[clip,width=0.95\columnwidth]{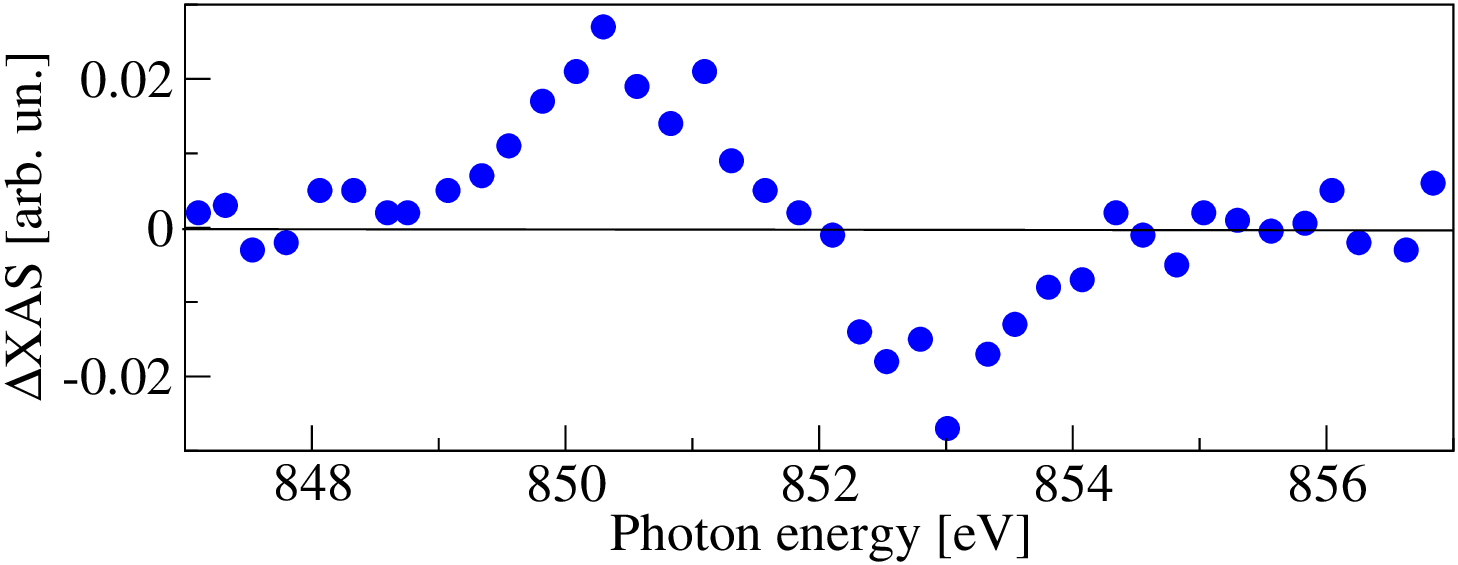}

\caption{\label{fig:XAS_sp_both} (Colour on-line) Top: 
The XAS $L_3$-edge spectra computed for Ni 
in equilibrium (black line) and non-equilibrium XAS spectra based on the theoretical
model with excitation density $f_{exc}=0.6$ (dashed red curve) and
assuming electron thermalization with $T_{e}=7000$\,K (dashed green
curve). The corresponding difference spectra between the two latter
and the equilibrium spectrum are shown in the middle panel. 
Bottom: Difference between experimental $L_{3}$
edge XAS spectrum of Ni 200 fs after the onset 
 of the pump laser and the XAS for equilibrium \cite{stamm07}.
 }
\end{figure}

Fig.~\ref{fig:XAS_sp_both}(top) shows the computed
equilibrium XAS spectrum at the Ni $L_3$-edge  as well as the computed 
non-equilibrium XAS spectra due to state-blocking and thermalization effects.
Both state-blocking and thermalization cause a shift of the calculated $L_{3}$ XAS peak
to lower energies and also small broadening of the peak width. This is
the very same modification of the XAS as the one that has been
observed experimentally \cite{stamm07}.
The XAS difference signal of the non-equilibrium and equilibrium spectra
exhibits a 5 eV wide,
oscillator-type structure centered around 852 eV which is very similar
to the one observed experimentally \cite{stamm07}, Fig.~\ref{fig:XAS_sp_both}(bottom).
The modification of the XAS spectrum can be understood from the depletion
and repopulation: first, down to 1.59 eV below $E_{{\rm F}}$ the
band states are depleted by the pump laser pulse. These states can
be accessed in the subsequent x-ray transition, which gives rise to
extra XAS intensity at lower energies. Repopulation removes furthermore
accessible states above $E_{{\rm F}}$, leading to a reduced intensity
at higher energies. Due to lifetime broadening the effects are distributed
over a several eV wide spectral range. %

\begin{figure}[tb]
\includegraphics[clip,width=1.0\columnwidth]{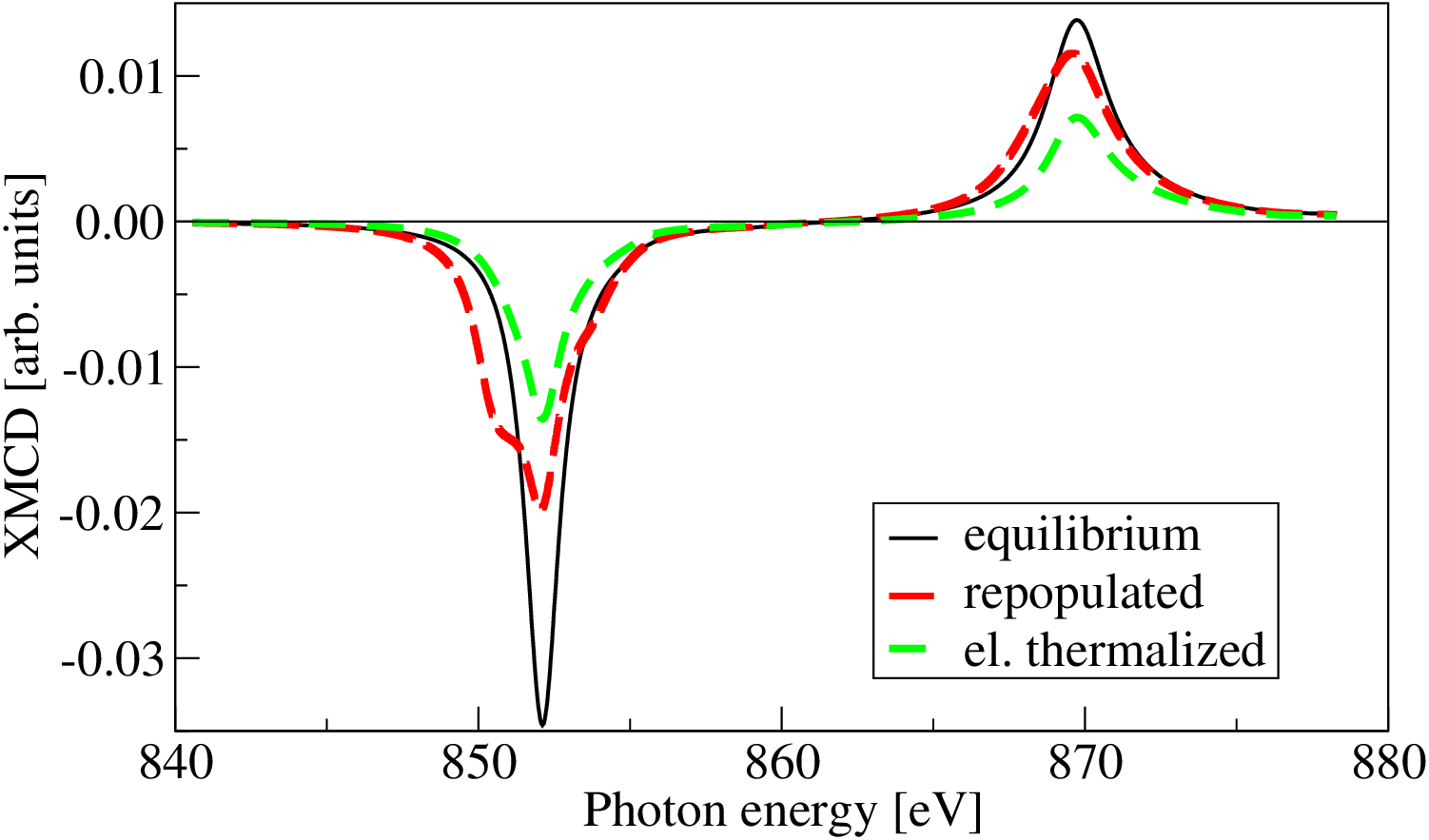}

\caption{ \label{fig:XMCD_spec_both} (Colour on-line) Calculated XMCD spectra
of Ni. Black line: equilibrium XMCD, red dashed line, with state-blocking
for an excitation density $f_{exc}=0.6$, and green dashed line, with thermalized
electrons of $T_{e}=7000$\,K.}

\end{figure}

Fig.\ \ref{fig:XMCD_spec_both} shows the full XMCD spectrum computed
with state-blocking or thermalization effects included. 
State-blocking for $f_{exc}=0.6$ reduces strongly the height of the
$L_{3}$ and $L_{2}$ peaks and increases the XMCD signal at the low-energy
sides of both $L_{3}$ and $L_{2}$ peaks. Thermalization also reduces
the XMCD peak at both $L$ edges, but does not create extra signal
at the low-energy sides. Note that in both cases the
change of the XMCD peak position is almost negligible when compared
 to the change in the XAS peak position (Fig.\ \ref{fig:XAS_sp_both}).

\begin{figure}[tb]
\includegraphics[clip,width=0.95\columnwidth]{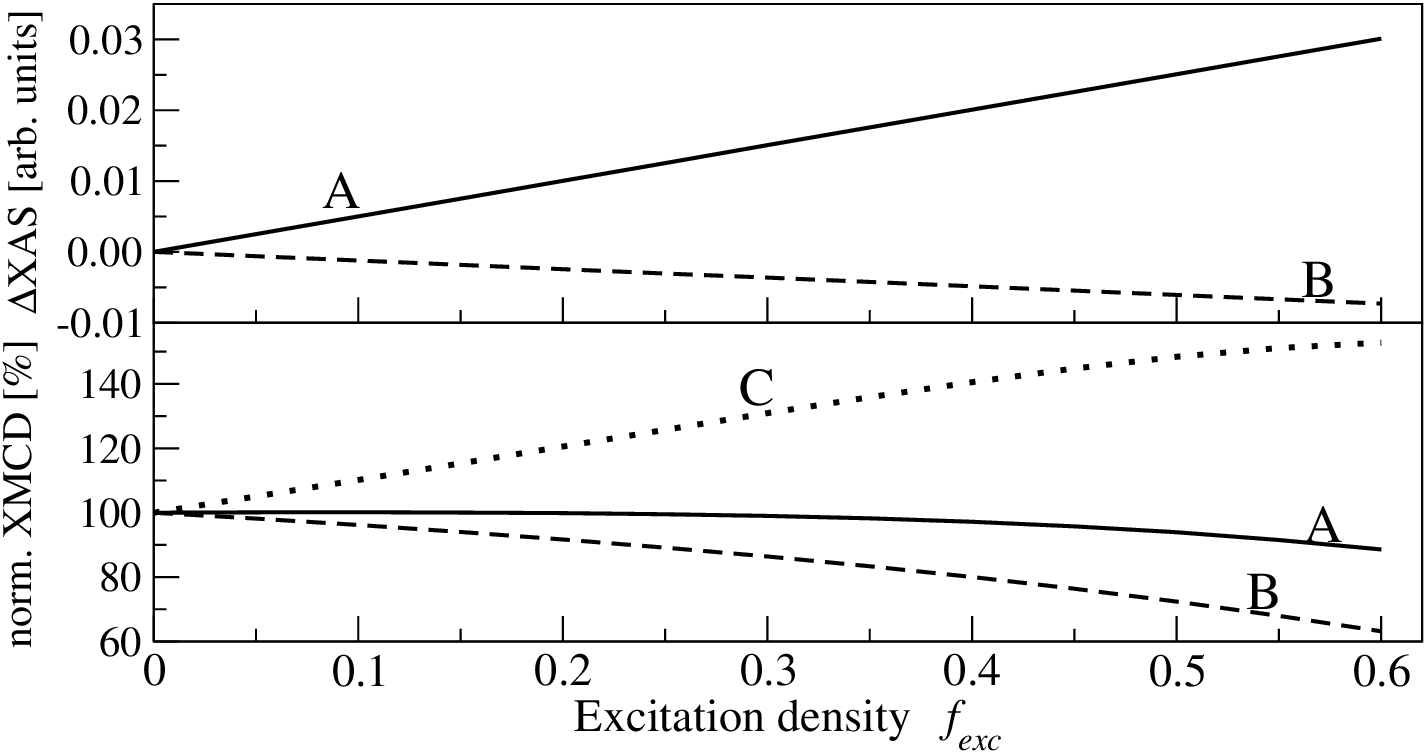}
%\medskip{}
\includegraphics[clip,width=0.99\columnwidth]{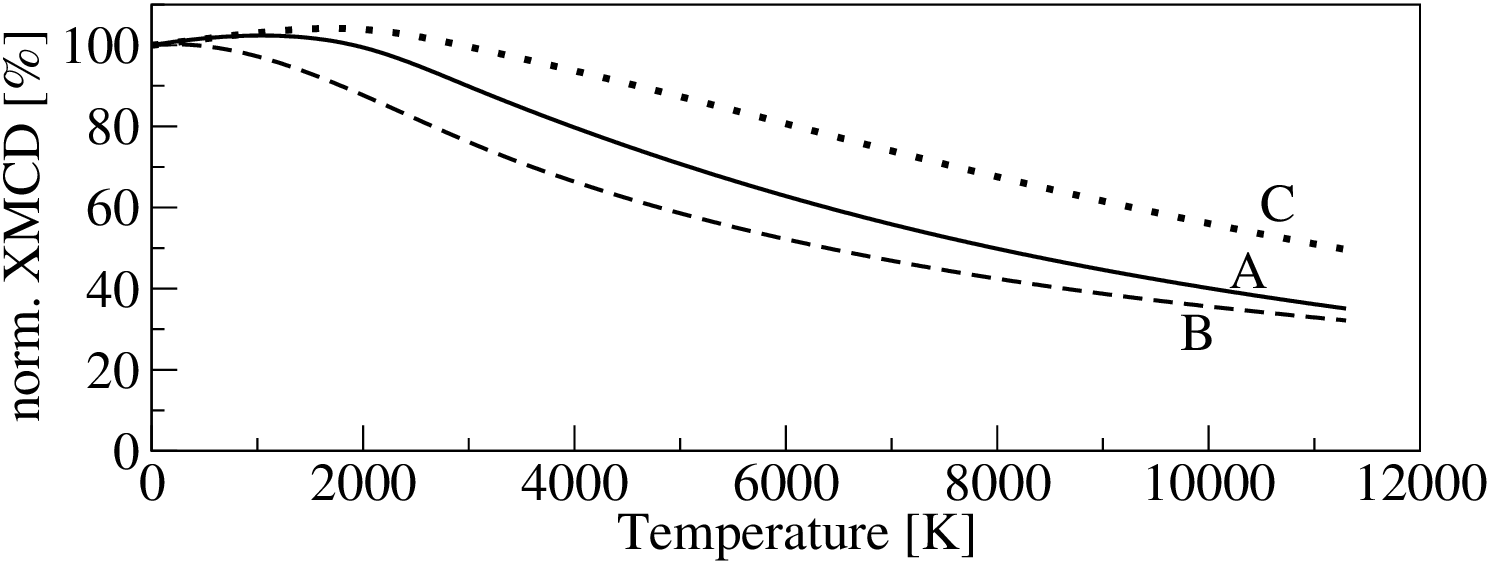}

\caption{ \label{fig:XAS_XMCD-fixEn} Change of the XAS and XMCD signals (normalized
with respect to its value without laser pulse) at the energies A,
B (and C for XMCD) as function of the excitation density $f_{exc}$
(top) and electron temperature $T_{e}$ (bottom).}

\end{figure}

To provide a detailed analysis, we study the response
at particular energies, denoted A and B in Fig.\ \ref{fig:XAS_sp_both},
which correspond 
to the XMCD maximum (B, at 852.15 eV) and to the XAS value equal to
one half of the XAS maximum (A, at 851.52 eV in our model) in
accordance with the experiment \cite{stamm07}. For later discussion
we introduce a third energy C given by the maximum of the XAS difference
curve (at 851.09 eV). The experiment \cite{stamm07} revealed a strong
reduction of the XMCD signal at energy B. % by about 65\%. 
The computed changes of XAS and XMCD at energy positions A, B, and
C depending on the excitation density $f_{exc}$ and temperature $T_{e}$
are shown in Fig.\ \ref{fig:XAS_XMCD-fixEn}. The XAS at A is growing
with the excitation density $f_{exc}$, while the XMCD at B is decreasing,
in agreement with the experiment. The change of these properties is
roughly linear in the excitation density (examined up to the value
$f_{exc}=0.6$), as was suggested earlier \cite{koopmans00}. In contrast
to the state-blocking, electron thermalization leads to an XMCD \textit{reduction}
over the whole edge. This happens because electron equilibration involves 
both a magnetization reduction and some blocking of optical transitions due to the broad FD 
distribution.
Noteworthy, it is possible to find such $f_{exc}$
and $T_{e}$ that the calculated relative XMCD reduction at B matches
the experimental value. 

A complete XMCD spectrum allows to obtain the total spin and orbital
momentum contributions to the magnetization via the XMCD sum rules
\cite{thole92,carra93}. The LSDA spin and orbital moments calculated
for Ni in the ground state are $m_{S}$=$0.62\,\mu_{\mathrm{B}}$
and $m_{L}$=$0.043\,\mu_{\mathrm{B}}$. Applying the sum rules (with
the usual neglect of the magnetic dipole term), we obtain $m_{S}$=$0.46\,\mu_{\mathrm{B}}$
and $m_{L}$=$0.042\,\mu_{\mathrm{B}}$. Thus, the orbital moment
is properly given, but the accuracy
for the Ni $3d$ spin moment is weaker; an error of up to 36\% with
respect to the exact value was reported previously \cite{wu94}. 
Applying the sum rules to the electron distribution for $f_{exc}=0.6$,
we obtain that $m_{S}$ changes from $0.46$ to $0.44\,\mu_{\mathrm{B}}$
and $m_{L}$ from $0.042$ to $0.037\,\mu_{\mathrm{B}}$. This finding
agrees (within the sum rules' accuracy) with the fact that the electron's
spin is preserved in the pump laser excitation, but the orbital moment
shrinks because transitions from $d$ to $p$ states are prevalent.
The XMCD sum rules are therefore valid for the studied excited system
with the same accuracy as for the ground state, but one must take
care when the perturbation of the XMCD spectrum is not homogeneously
distributed over the edges. 
Although the XMCD signal at the $L_{3}$ peak is reduced $1.6$ times
in our simulation, the area of the whole $L_{3}$ edge is reduced
only 5\%, 
because of the compensating XMCD signal rise near energy C. Consequently,
taking the XMCD peak at B as a measure for the reduction of the spin
moment would unjustly suggest a significant demagnetization, which
would not be observed when the whole spectral area is considered.

Also in the case of electron thermalization, we find--from an analogous
comparison--that the sum rules remain valid. Replacing the spectral
integral of the sum rules with the XMCD value at B is in this case
a better approximation, as the XMCD signal over the whole edge is
reduced. However the reduction is not homogeneous. If we consider,
for example, a $T_{e}$ of 5000 K, a reduction of the XMCD at B of
42\% is computed, but the reduction at C is only 12\% (Fig.\ \ref{fig:XAS_XMCD-fixEn}).
This difference is due to a partial blocking of states that results
from the high-$T_{e}$ FD function. A direct calculation of the moment in
the thermalized state gives $m_{S}$=0.39 $\mu_{B}$ ($m_{L}$=0.024
$\mu_{B}$) which implies a reduction of the spin moment by 37\%.
An application of the integral sum rules gives a reduction by 33\%,
in reasonable agreement.

\section{Discussion of the temporal evolution} %new text
In the case of state-blocking a linear increase of the XAS at A with
$f_{exc}$ is predicted, which is due to the shift of the $L_{3}$
peak to lower energy. A smaller increase of the XAS at A follows
from electron thermalization with a FD function having
a high $T_{e}$ (see Fig.~\ref{fig:XAS_sp_both}).
Conversely, the XMCD signal at its maximum (point B) is reduced for both cases of 
state-blocking and high electron temperature, but for the XMCD peak value the latter effect is
slightly stronger, i.e. the XMCD peak maximum is reduced less by the state-blocking effect than by the
corresponding high-$T_{e}$ state (Fig.\ \ref{fig:XMCD_spec_both}). This observation
allows us to predict that the crossover from the non-equilibrium, state-blocking state to the electron-thermalized state
should be accompanied by a {\it reverse} of the initially increasing trend of the XAS (at A) 
and, contrary to it, an almost unchanged or slowly decreasing trend of the XMCD maximum at B. 
At larger times a slow increase of the XMCD at B is, however,  expected, due to the electron-lattice equilibration, which acts as a competing factor.  

The experiment \cite{stamm07} shows a linear
rise of the XAS signal at A with time up to about 220 fs after the onset of the 70 fs pump
pulse, after which the XAS starts to decrease.
 A \textit{different} temporal behavior is observed 
for the XMCD at B: the XMCD signal reduces linearly with time up to
220 fs after the onset of the pulse, after which it remains roughly
constant. This indeed confirms the behavior  we predicted above for the crossover to 
electron thermalization. The here presented
theory asserts therefore  that for Ni in the studied experiment \cite{stamm07}, 
state-blocking is prevalent up to about 150 fs after the 70 fs pulse, when
the crossover to electron thermalization has occurred, which enforces a reduction of the
XAS at A.  
We note that the here deduced relaxation time of 220 fs for the crossover to electron thermalization in 
Ni is consistent with earlier pump-probe magneto-optical measurements \cite{koopmans00}, whereas a
moderately smaller relaxation time of 150 fs after the onset of the pulse
was measured on CoPt$_3$ films \cite{r_02_Guido_CoPt3_MOValidity}.

\section{Conclusions}
In Ref.\ \cite{stamm07} important conclusions regarding the spin
and orbital moment evolution were drawn through application of the
XMCD sum rules. It is essential to note, however, that the change
of the spectral integral over the whole $L_{3}$ edge was estimated
from the XMCD value at only one energy (at point B). From the presented calculations
we conclude that the change of the spectral integral
over the whole $L_{3}$ edge cannot be estimated from the XMCD value
at only one energy (B) if state-blocking is present. It is therefore
not possible to decide about the role of transfer of spin angular momentum
to the lattice from the present experimental data \cite{stamm07}, missing
a complete femtosecond-resolved XMCD spectrum. 

As we have found that the XMCD sum rules (in their integral form) are valid
even in the laser excited state, a complete XMCD spectrum of laser-pumped
Ni should thus provide reliable information on the evolution of spin
and orbital moments. We demonstrate that state-blocking and
high electron temperature effects lead to the following changes in the spectra: the shift of the XAS
 to lower energy, the shape of the XAS difference spectrum, the rise of the XAS at A, and
the strong reduction of the XMCD peak at B in the first 220 fs after the onset of the pulse.  
This implies that the presence of non-equilibrium electron distributions can largely explain 
the experimental observations
\cite{stamm07} without the suggested transfer of spin angular momentum to the lattice.
Our calculations also predict the approximate time evolution
of these changes related to the crossover to thermalization, which we estimate to occur at about 150 fs after ending of
the pump pulse.
We propose that other
experimental data, especially those obtained for femtosecond-resolved XMCD at energies
near position C, together with the presented theory, would allow to
separate unambiguously the various contributions to the
apparent XMCD and pave the way for a precise understanding of 
the nature of the femtosecond demagnetization as well as
 reveal the timescales crucial for the demagnetization process.

\acknowledgments

We thank %B. Koopmans and 
H. D{\"u}rr, C. Stamm and J. Rusz for valuable discussions. Support through the Swedish
Research Council (VR), STINT, INTAS, the G. Gustafsson Foundation,
the European Community's Seventh
Framework Programme (FP7/2007-2013) under grant agreement No.\ 214810, 
{}``FANTOMAS\char`\"{}, 
and the Swedish National Infrastructure for Computing (SNIC) is acknowledged. 

%\vspace{-0.5cm}
%\bibliographystyle{/home/karel/latex/revtex4/apsrev}
%\bibliography{/home/karel/latex/b_kc}
%\bibliographystyle{eplbib}
%\bibliography{/home/karel/latex/b_kc}

%\end{document}

\end{document}